\documentclass[journal=jacsat,manuscript=article]{achemso}

\usepackage[version=3]{mhchem} 

\author{Sanjog V. Joshi}

\author{Pradeep R. Nair}

\email{sanjogvjoshi@gmail.com; prnair@ee.iitb.ac.in}

\affiliation[Indian Institute of Technology Bombay]
{Department of Electrical Engineering, Indian Institute of Technology Bombay, Mumbai India}

\title[An \textsf{achemso} demo]
  {Electronic Circuit Inspired Optimization of Nanogap Electrochemical Biosensors}

\keywords{Biosensor, reaction-diffusion, detection limits,
biochemical reaction, Electrochemical Redox cycling scheme
}

\begin{document}

\begin{abstract}
  Electrochemical biosensors and the related concept of redox detection at nanogap electrodes are increasingly explored for ultra-sensitive detection of biomolecules. While experimental demonstrations have been encouraging, the associated design and optimization of electrode geometry, beyond the simple one-dimensional architectures, is inherently challenging from multiple aspects related to numerical complexity. Here we develop a facile simulation scheme to address this challenge using well established electronic circuit analysis techniques that are available as open source-ware. Based on this approach, we show that electrode geometry, especially nano-structured redox electrodes on a planar surface, has interesting implications on the detection limits and settling time of electrochemical biosensors. The methodology we developed and the insights obtained could be useful for electrode optimization for a wide variety of problems ranging from biosensors to electrochemical storage.
\end{abstract}

\section{Introduction}
Biosensors,
which aim at detection/measurement of the concentration
of molecules in samples \cite{C3CS35528D} find wide-ranging applications like diagnosis (blood, glucose, proteins, tumors, heart seizure risk,
etc.) \cite{Song2006}, pollution and contamination detection in
various environments (water, tissues, air, food, etc.) \cite{doi:10.1002/jcb.26030},
in the pharmaceutics industry (drug discovery and analysis,
chemical compound dosing, clinical validations, etc.) \cite{doi:10.1080/00032710500205659,Turdean,10.1371/journal.pone.0026846}, fundamental research in biology (cell signaling
measurement, DNA sequencing, cell detection, metabolic
Engineering) \cite{Label,VERCOUTERE2002816,C7LC01117B,Yan2016}, etc.
Of the many competing technologies, electrochemical detection \cite{B714449K} based on redox cycling has reported significant recent progress \cite{doi:10.1021/nn404440v,doi:10.1002/adfm.201907701}.
\vspace{0.7em}
Among the electrochemical detection strategies, single electrode-based detection schemes like chronoamperometry has inherent disadvantages related to diffusion-limited response \cite{BAUR2007829}. Further, a target molecule or the associated mediator results in a single event of charge transfer at the electrode. Nanogap based electrochemical detection circumvents this limitation through two or more electrodes configured in a redox detection mode \cite{ZAFARANI201642}. Here successive oxidation and reduction of analyte molecules at two closely spaced electrodes lead to current amplification and, thus, the detection limit of electrochemical sensors is significantly better \cite{doi:10.1021/ac00124a045}. Despite this interesting concept and significant experimental results, the design and optimization of electrode geometries for such a sensor is challenging due to the inherent numerical complexities. \vspace{0.7em} 

In this manuscript, we develop a scheme to explore the optimization of redox detection at nanogap electrodes. As such, this is a complex challenge related to numerical simulations as one needs to account for diffusion of two species along with non-linear Butler-Volmer reaction kinetics at the electrodes. However, we show that this can be conveniently addressed based on electronic circuit simulation tools \cite{10.1371/journal.pone.0182385}.
Further, scaling to complex geometries is simple and evident - both conceptually and implementation wise. Based on this approach, we show that detection limits and settling times have unique scaling trends with nano-structuring of electrodes. Below, we first describe the methodology and then address the optimization aspects. 



\section{Model System and Methodology}

\begin{figure}[h!]
\center
\includegraphics[scale=1]{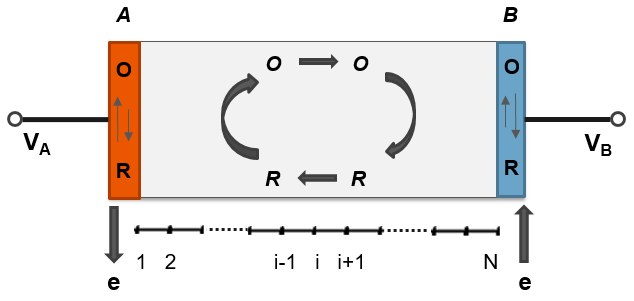}
\caption{A generalized scheme for nanogap based redox detection. Also shown is a grid over which the transport equations are discretized for numerical solution.
}
\label{Electrode_Geometries}
\end{figure}

A generalized scheme for nanogap based redox detection is shown in Fig. \ref{Electrode_Geometries}. Here, the two electrodes are so configured that a redox species can shuttle between them, which results in an amplified electrode current. For example, at electrode A ($E_A$), the species R gets converted to O, which then diffuses to electrode B ($E_B$) where it gets converted back to R again. So every additional molecule of R (or O) results in multiple charge transfer reactions at both electrodes, thus increasing the current significantly. 

The time dependent electrode currents can be obtained through the simultaneous solution of transport of redox species as well as the electron transfer reactions at the electrodes. Under the assumption of negligible electric field or fluid motion, the transport of redox species is given by \\
\begin{equation}
 \frac{\partial \rho_{R,O}}{\partial t} =D_{R,O}{ \nabla^2}{\rho_{R,O}},
\end{equation}

where the term $\rho$ denotes the density of either R or O species (denoted by the subscript), and D denotes the corresponding diffusion coefficient. The electrode current density due to electron transfer reactions at the electrodes A (at $x=0$) and B (at $x=W$, W is the spacing between the electrodes or the nanogap) is given as:
\begin{multline}
 J(x=0)=nF(K_{FA}\rho_{R}(x=0)-K_{RA}\rho_{O}(x=0))
\\
 J(x=W)=nF(K_{FB}\rho_{R}(x=W)-K_{RB}\rho_{O}(x=W))  
\end{multline}

Here, $K_{F}$ denote the $R\rightarrow O$ reaction constant while $K_{R}$ denotes the $O\rightarrow R$ reaction constant. The subscript A or B denotes electrode A or B, respectively. $F$ is the Faraday's constant, and $n$ denotes the electrons involved in a single reaction. In general, the reaction rates depend on the potential applied at the electrode and the formal potential of the species and is represented in the Butler-Volmer formalism \cite{Bard1980ElectrochemicalMF}
(i.e., $K_{F} \propto e^{q(V-V_F)/kT}$, where V is the applied potential, and $V_F$ is the formal potential of the reaction). 

\vspace{0.7em}
Optimization of the electrode geometry for ultrasensitive detection of biomolecules is not an easy task. For the simple 1D cases, compact analytical results are available for the electrode current under the diffusion-limited regime. However, similar analytical results are not available for the optimization of more complex electrode geometries, which require detailed numerical simulations. Given the non-linear boundary conditions imposed by Eq. 2, access to accurate numerical simulation packages is also not readily available to the experimental community. In this regard, below, we first describe one such scheme and show its implementation through electronic circuit analysis - much like the well-known example of solving transport problems in biology through an analogy to electrostatics\cite{BERG1977193}. 
\vspace{0.7em}
Of the several techniques to solve the above equations numerically, the finite difference (FD) scheme involves converting the partial differential equations over the region of interest into a set of algebraic equations over a set of grid points and then numerically solving them \cite{book}.
One such spatial grid is shown in Fig. \ref{Electrode_Geometries}, where the grid point or node 1 corresponds to electrode A and node N corresponds to electrode B. For simplicity, a uniform grid is considered (which need not be the case) which results in $\triangle x=W/(N-1)$. For a $1D$ case, an altered form of FD representation of eq. (1) (done purposefully to develop an equivalence with electronic circuits, as will be evident later) is

\begin{multline}
\triangle x\frac{\partial \rho_R(i)}{\partial t}=\frac{D_R}{\triangle x}(\rho_R(i+1)-\rho_R(i))
+\frac{D_R}{\triangle x}(\rho_R(i-1)-\rho_R(i)),
\end{multline}

where the variable $i$ denotes the location (i.e., a particular grid point). The above equation can be interpreted in simple terms as: The LHS denotes the rate of increase in the concentration of R at location $i$, over a grid spacing of $\triangle x$. This is nothing but the net diffusion flux towards it from the neighboring nodes (as shown in RHS, with the flux being proportional to the concentration difference between the node points).

\vspace{0.7em}
While the above is sufficient to account for the transport of R (similar treatment needed for O as well), the equivalent FD representation at the electrode is more complex. At the first node (i.e., $x=0$), one needs to account for three aspects - (i) the change in concentration, (ii) the diffusion flux, and (iii) the reaction happening at the electrode and its contribution to the concentration change. Application of detailed balance (which is also known as continuity equations in semiconductor theory \cite{pierret1996semiconductor}), indicates that the appropriate FD representation of the reaction and diffusion processes at the electrode interface, with $i=1$, is given by:

 \begin{multline}
  \frac{\triangle x}{2}\frac{\partial \rho_R(i)}{\partial t}=\frac{D_R}{\triangle x}(\rho_R(i+1)-\rho_R(i))
  +(K_{FA}\rho_{R}(i)-K_{RA}\rho_{O}(i))   
 \end{multline}

Similar equation at $x=W$ (or $i=N$) holds good for the reaction and diffusion process at electrode B. Equations (3) and (4), appropriately set up at every node for both R and O species, can be numerically solved to obtain the transient as well the steady-state electrode current in a nanogap based redox detection scheme. While it could certainly be attempted in a conventional manner by setting up matrix equations and backward Euler scheme, a direct analogy to circuit analysis (i.e., Kirchoff's laws) provides an opportunity to solve the same complex problem using well established EDA tools- thus simplifying the optimization efforts.

\vspace{0.7em}
As already mentioned, there exist a well-known analogy between electrostatics and diffusion problems in biology \cite{BERG1977193}. Specifically, under steady-state conditions, diffusion equation, and Laplace equation of electrostatics have the same functional form. Hence the parameter pairs - voltage and density, charge and diffusion flux becomes equivalent, and thus solutions/insights in one field can be easily adapted to others with appropriate change in parameters. There exist a similar analogy to circuits as well as the current through a resistor depends on the potential difference between its nodes (i.e., here the equivalent parameters are potential and concentration, electric current and diffusion flux, etc.). So, the equation (3) can be represented by an equivalent circuit where the node i is connected to its neighboring nodes with the resistors of value $R_t=\triangle x/D$, subscript t denoting the transport. While this accounts for the RHS, the LHS indicates that every node should be connected to the ground through a capacitor of value $C_i=\triangle x$ as well, subscript i denoting the internal node. 

\vspace{0.7em}
The treatment for the boundary nodes, where electrode reactions need to be accounted for, is different from that of the internal nodes. A close scrutiny of eq. (4) indicates that, for node 1, in addition to a resistor which connects it to node 2, there should be an additional resistor $R=1/K_{FA}$ connected to the ground and an additional dependent current source $I=K_{RA}\rho_{O}$ as well.  Note that this current source depends on the concentration of O at $x=0$. Further, the LHS indicates that a capacitor $C_b=\triangle x/2$ should be connected to the ground, subscript b denoting the boundary node. 

\vspace{0.7em}
Once the above described conceptual mapping to circuit elements is done, transient diffusion problems could be easily solved using EDA tools, thus reducing the computational complexity. We note that a similar approach was reported earlier \cite{8602461} to solve transient diffusion problems of single species for ISFET sensors. However, they used different formalism for R and C ($R=\triangle x^2/D$, and $C=1$, which results in the same $RC $ product as our proposed scheme). Such a formalism is not equipped to incorporate electrode kinetics in a nanogap sensor. Our methodology, described above, is conceptually correct and dimensionally consistent and hence is well suited to address a multitude of scenarios related to electrochemical sensors, as illustrated in subsequent sections.

\section{Modeling of Nanogap sensors}
We first illustrate the capability of our formalism by addressing the ID system of nanogap sensors, after which the optimization of electrode geometry is attempted.
\vspace{0.7em}

\begin{figure}[h!]
\center
\includegraphics[scale=0.6]{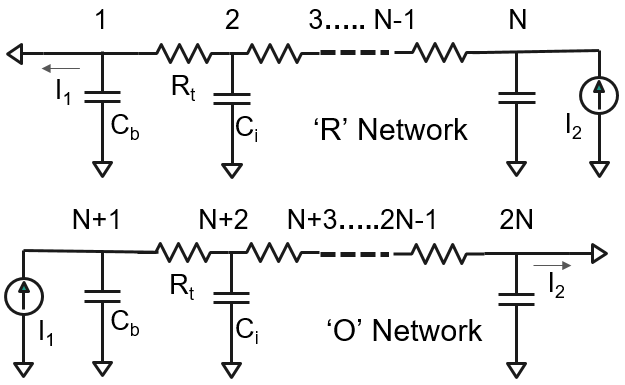}
\caption{Equivalent  circuit for nanogap based redox detecton in diffusion-limited regime.
}
\label{Nanogap1}
\end{figure}

\textbf{ID system: Diffusion limited response}
For the diffusion-limited case, we need to solve the transient diffusion equation for both R and O subject to the condition that the concentration of R at electrode A is zero, while the concentration of O at electrode B is zero. Further, the rate of consumption of R at electrode A is the same as the rate of generation of O at electrode A (and similar condition at electrode B). This requires construction of not just one, but two equivalent circuits (each representing the transport of only one species) with the dependent current sources as shown in Fig. \ref{Nanogap1}. Note that the node with number $i$ in the first circuit represent the same spatial location as the node $N+i$ in the second circuit. Accordingly, the current that sinks at node 1 is the electrode current due to $R\rightarrow O$ reaction at electrode A which is the same current that sources the generation of O in the second circuit at node $N+1$. A similar argument explains the electrode current and circuit elements at electrode B.

\begin{figure}[h!]
\includegraphics[scale=0.4]{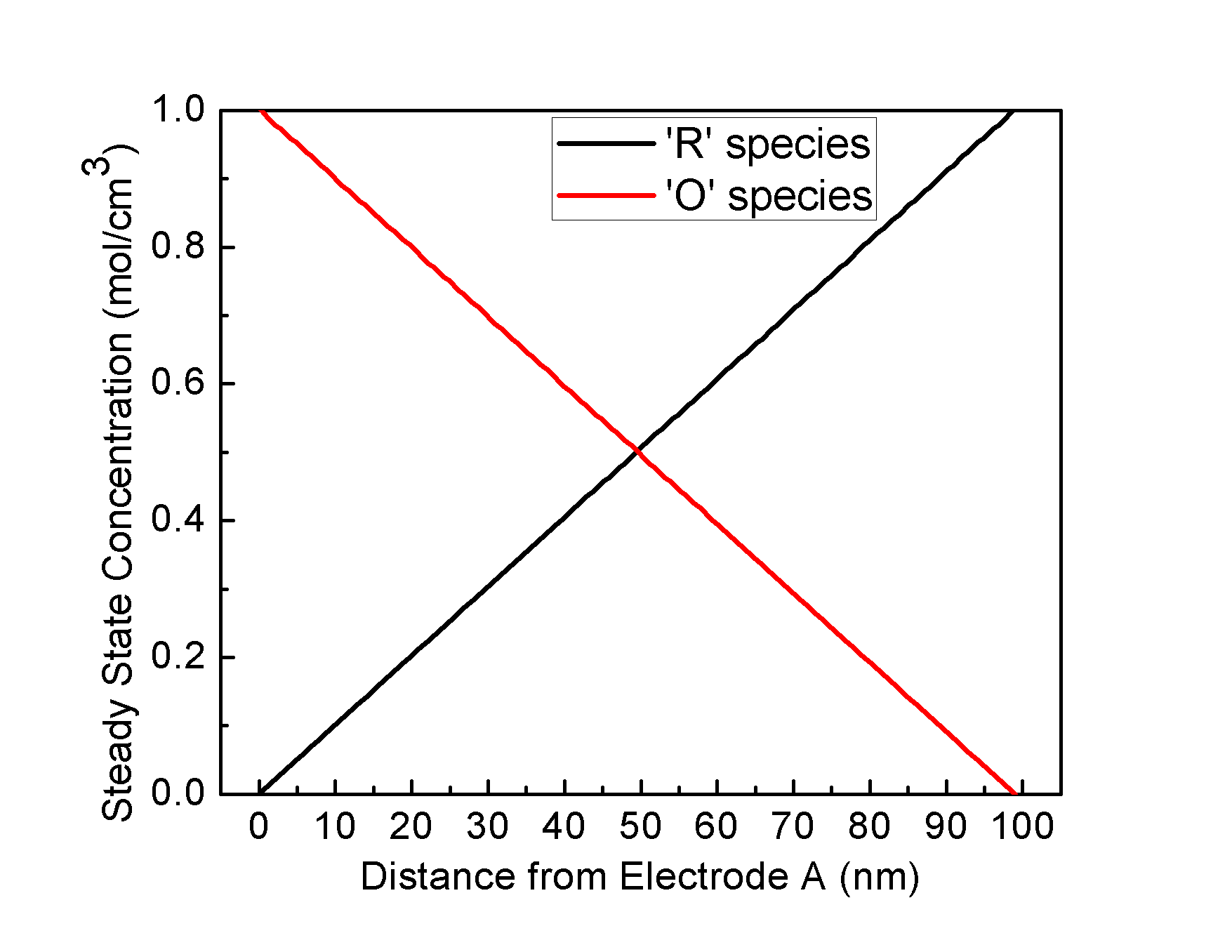}
\caption{Normalized steady state concentration profiles of the redox species in a nanogap sensor with $W=100nm$}.
\label{Nanogap_Concentration}
\end{figure}

\begin{figure}[h!]
\includegraphics[scale=0.4]{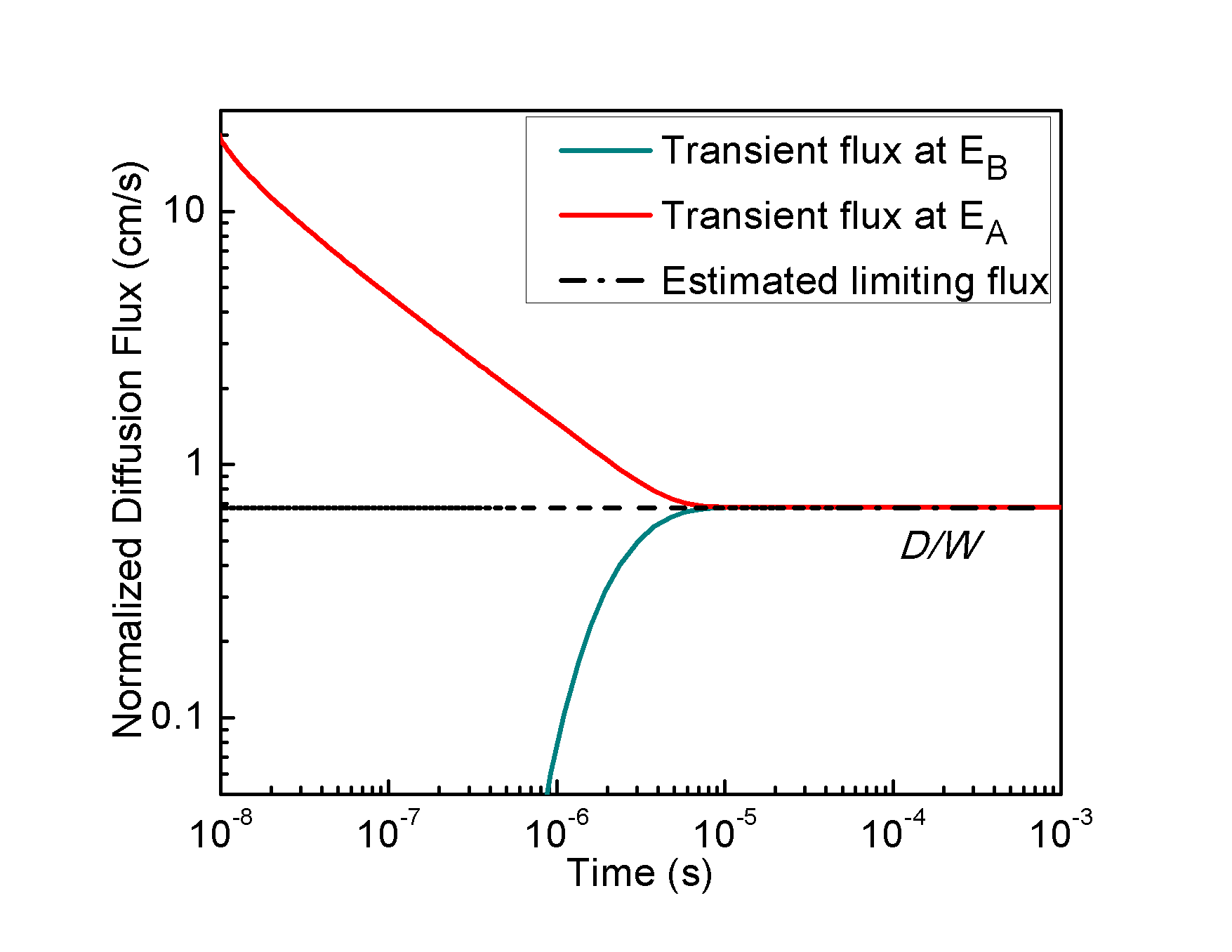}
\caption{Transient diffusion flux per unit concentration at the electrodes of nanogap sensors in a diffusion-limited case. 
}
\label{Current_Amplification}
\end{figure}

The above system of coupled circuits can now be used to explore the performance of nanogap sensors (with $W=100nm$, see Fig. \ref{Electrode_Geometries} for schematic). Figure \ref{Nanogap_Concentration} indicates that the concentration profiles are linear under steady-state conditions as predicted by the diffusion equation. The transient features are shown in Fig. \ref{Current_Amplification}. The initial condition is such that only R species is present (density of $\rho_R$) at $t=0$. Accordingly, the initial current at electrode A is several orders of magnitude larger than that at electrode B. Further, the initial decay of the transient current at electrode A shows the well known $t^{-0.5}$ dependence (Cottrell current). It is also well known that the steady-state current density for
two planar electrodes \cite{ZAFARANI201642} is given by
\begin{equation}
 J=\frac{I}{A}=nF\frac{\rho_RD}{W},  
\end{equation}
where n is the number of electrons involved in the redox reaction, and $\rho_R$ is the initial analyte concentration. The results shown in Figs. \ref{Nanogap_Concentration}-\ref{Current_Amplification} compare very well with analytical predictions and hence validates the proposed methodology.

\begin{figure}[h!]
\center
\includegraphics[scale=0.6]{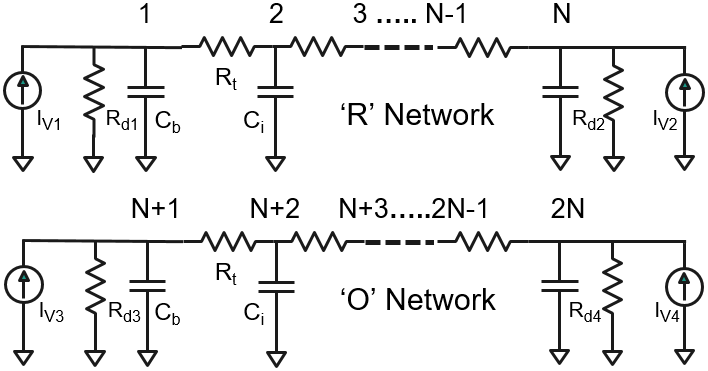}
\caption{Equivalent  circuit of the generalized scheme for the nanogap based redox detection in 1-D incorporating finite reaction rates at electrodes. $I_{V1}=K_{RA}V_{N+1}$, $R_{d1}=1/K_{FA}$, $I_{V2}=K_{RB}V_{2N}$, $R_{d2}=1/K_{FB}$, $I_{V3}=K_{FA}V_{1}$, $R_{d3}=1/K_{RA}$,  $I_{V4}=K_{FB}V_{N}$, $R_{d4}=1/K_{RB}$.   
}
\label{Nanogap2}
\end{figure}

\textbf{ID system: Generalized scheme}
Incorporation of finite reaction rates at electrodes as circuit elements is the next challenge to adapt EDA techniques for electrochemical sensors. However, this can be achieved in an insightful manner: Eq. (2) indicates that the current at any electrode depends on the concentration of R and O molecules near the surface. As far as the R network is concerned, the dependence on R is nothing but a resistive component ($R_{d1}=1/K_{FA}$, subscript d denoting degradation resistor). But the dependence on O, which is part of the second network, can be conveniently represented by a dependent current source ($I_{V1}=K_{RA} \rho_{O}$ or $I_{V1}=K_{RA}V_{N+1}$). A similar procedure is followed for electrode B in the R network and then for the electrodes in the O network. Based on these insights, the equivalent circuits, along with the values of resistors and current sources (in caption) for the generalized scheme is as shown in Fig. \ref{Nanogap2}. The convention regarding the equivalence of nodes, circuit elements, etc. remain the same as in the previous section.\\

\begin{figure}[h!]
\includegraphics[scale=0.4]{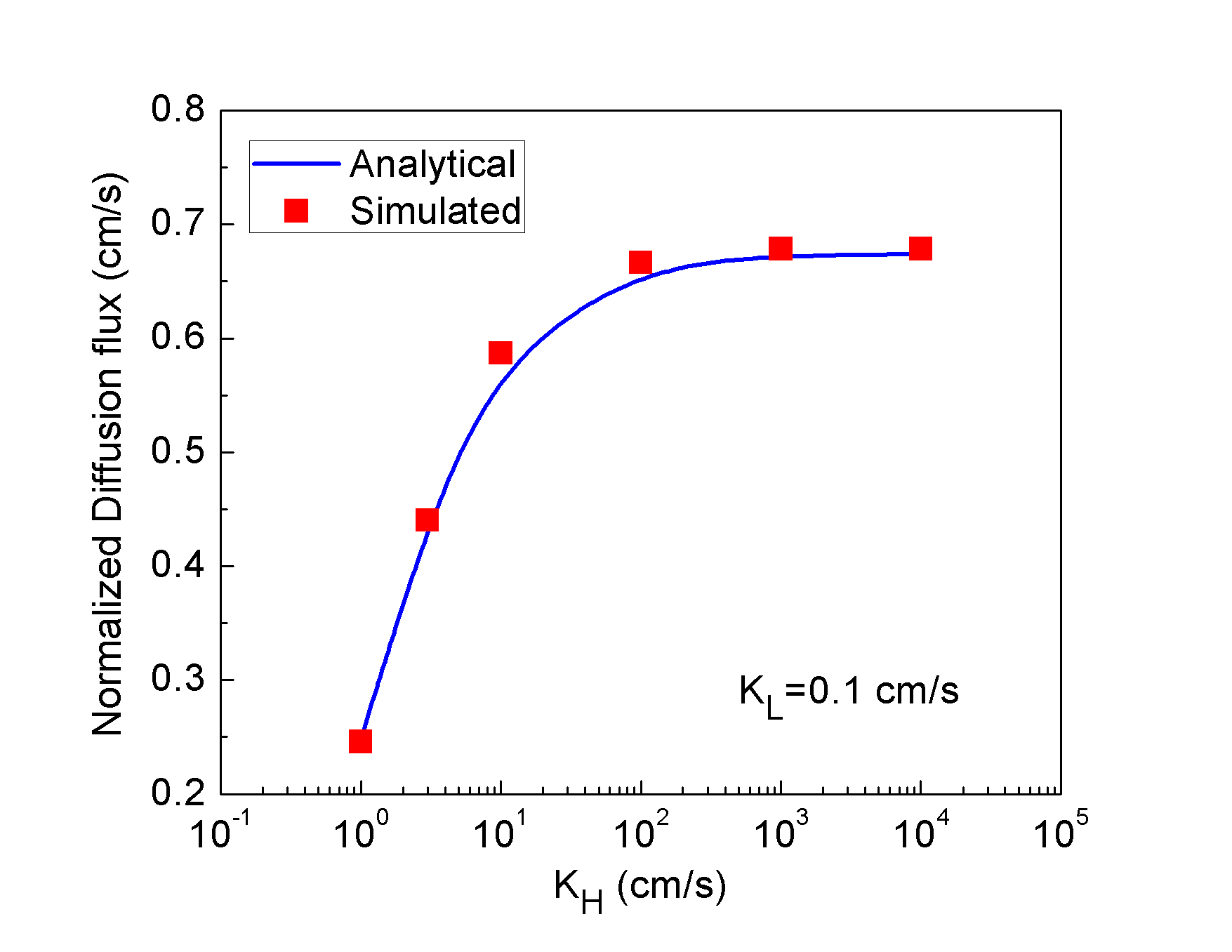}
\caption{Comparison of numerical simulation results with analytical predictions for steady-state diffusion flux per unit concentration in a nanogap sensor as a function of reaction rates.
}
\label{Reaction_Limited_1}
\end{figure}

Analytical expression for steady-state current density \cite{C2AN35346F} in the reaction-diffusion regime is
\begin{equation}
J=nF\frac{\frac{\rho_RD}{w}(\frac{K_H-K_L}{K_H+K_L})}{1+\frac{D}{w}( \frac{2}{K_H+K_L})} , 
\end{equation}

where $K_H$ and $K_L$ are the rate constants at the electrodes (subscripts H and L denote the higher and lower values of rate constants, respectively). Assuming that oxidation is the favorable reaction at electrode A and reduction is the favorable one at electrode B, we assign $K_{FA}=K_H, K_{RA}=K_L, K_{RB}=K_H$, and $K_{FB}=K_L$ (for convenience, same $K_H$ value is used for favorable reactions at corresponding electrodes and vice versa). Figure \ref{Reaction_Limited_1} shows a comparison of the numerical simulation results and analytical predictions. The generalized scheme predicts the nanogap sensor response over several orders of magnitude change in the reaction parameters.

\begin{figure}[h!]
\includegraphics[scale=0.4]{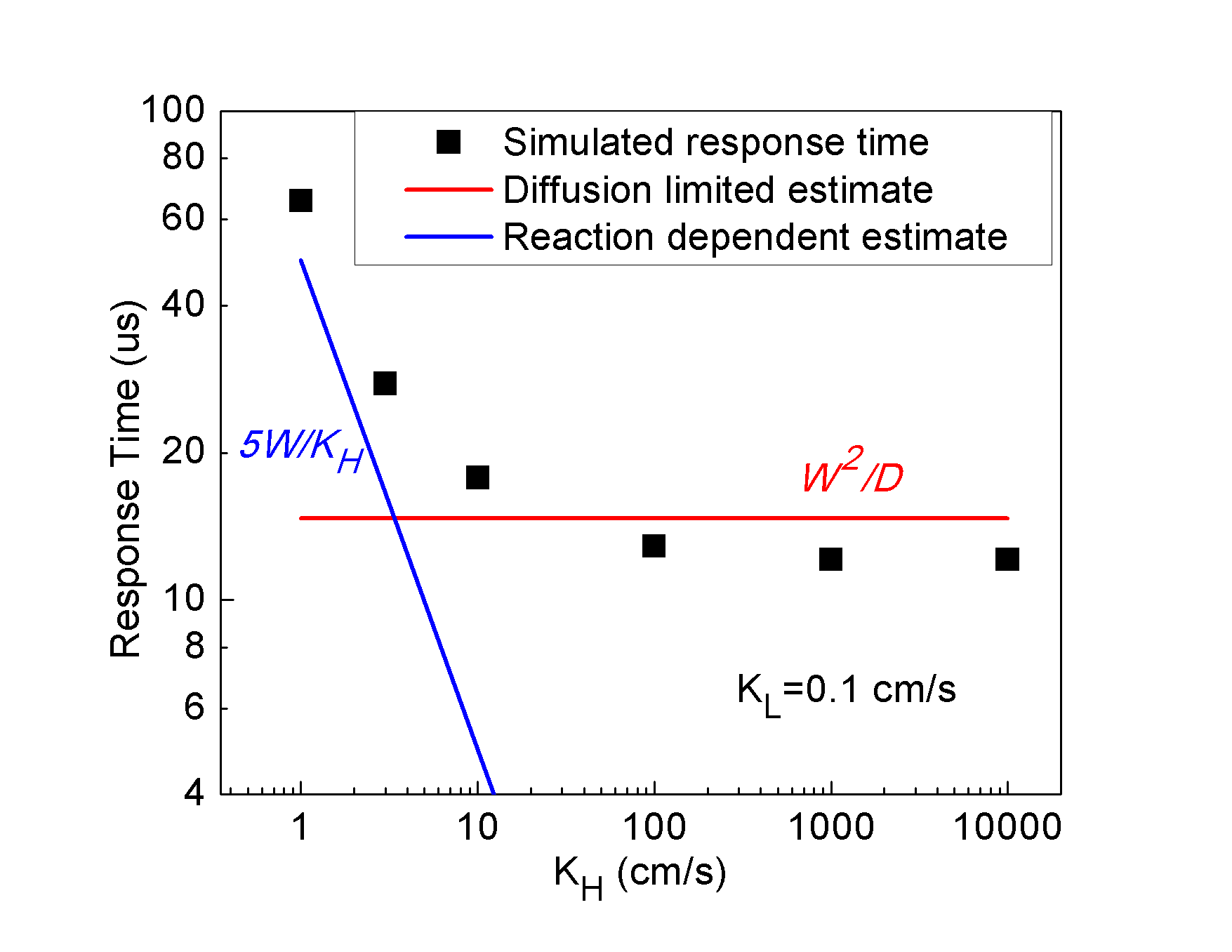}
\caption{Numerical simulation results for the settling time for a nanogap sensor. Two regimes of operation, reaction limited and diffusion-limited, are clearly evident along with analytical estimates for the settling times }
\label{Reaction_Limited_2}
\end{figure}

The time taken to reach steady-state conditions, or the settling time, is an important parameter for any sensor system and is plotted in Fig. \ref{Reaction_Limited_2} as a function of reaction rates. Also plotted is asymptotic analytical estimates for the same. Under the diffusion-limited regime, the time taken to reach a steady-state is nothing but the time taken by the molecules to shuttle between the two electrodes, which is given by $W^2/D$. For the reaction dominant case, the settling time is not established in the literature and can be obtained by a simple analysis. At any time, the number of molecules $\triangle N$ that undergoes reaction at the electrode surface (per unit area) over a duration $\triangle t$ is given as
\begin{equation}
 \triangle N= K_H \rho_R \triangle t   
\end{equation}

We know that the integrated number of molecules in the system (again, per unit area) at $t$, under such conditions (i.e., reaction limited with no transport limitations), is $N= \rho_RW$. Accordingly, the associated time constant is $\tau=W/K_H$, and the settling time could be 5$\tau$. Fig. \ref{Reaction_Limited_2} shows that the analytical estimates for the settling time compares well with the numerical simulation results. Interestingly, the same figure also provides a thumb rule for the transition from reaction limited to transport limited scenarios in terms of $D$, $W$, and $K_H$.

\section{Nano-structured Electrodes}
\begin{figure}[h!]
\includegraphics[scale=0.55]{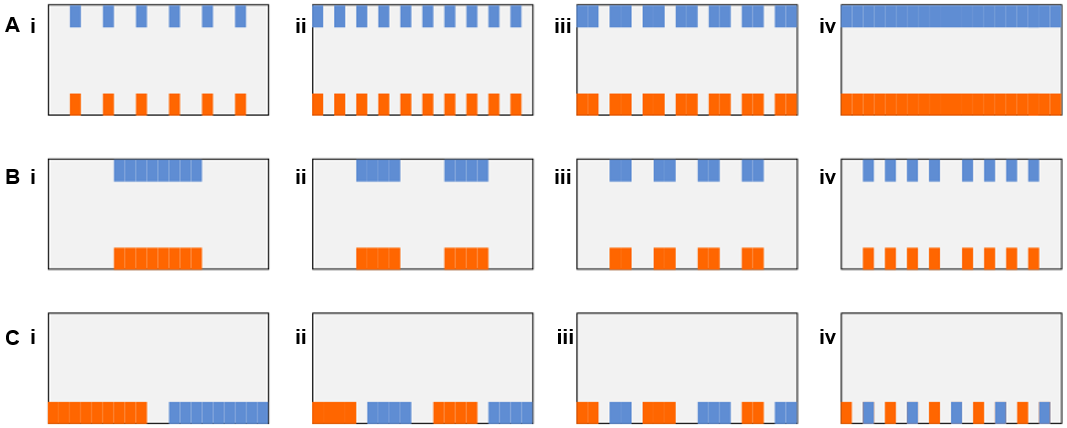}
\caption{Nanostructured electrode geometries for redox detection. A) Redox electrode array with increasing electrode area. B) Alternate redox electrode array scheme with the net electrode areas kept the same C) Nanostructured electrodes with planar layout. This simulation domain is $200nm \times 100nm$. The individual electrodes are essentially metal sheets with zero thickness, although shown with finite thickness. The color scheme for electrodes is the same as in Fig. 1.
}
\label{2D_Electrode_Geometries}
\end{figure}

Here we extend the generalized scheme developed in previous sections to explore complex nanostructured electrode geometries. The various schemes considered are shown in Fig. \ref{2D_Electrode_Geometries}. Here Scheme A corresponds to an array of redox electrodes whose net area (or density) increases from left to right. Indeed, the structure listed as $iv$ is nothing but a system of parallel electrodes. In B, an alternate scheme is explored where the electrode area is kept the same, but the layout is changed. Contrary to schemes A and B, a planar layout is attempted in scheme C. Note that for all the structures shown in Fig. \ref{2D_Electrode_Geometries} the simulation domain is the same: $200nm \times 100nm$. Also, the minimum separation between two consecutive electrodes in these structures is $10nm$. The simulations are performed in 2D geometry through an extension of the 1D discretization discussed earlier. To explore the effects of nanostructured electrodes, we restrict our analysis to the diffusion-limited regime only.

\vspace{0.7em}

\begin{figure}[h!]
\includegraphics[scale=1]{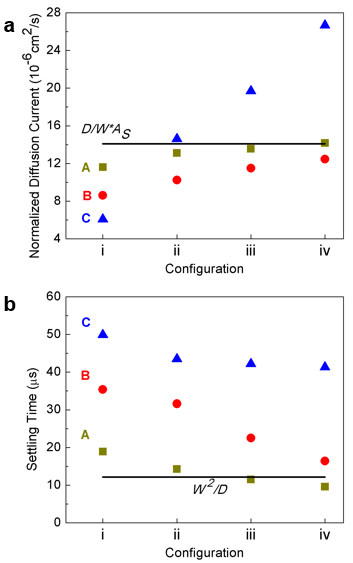}
\caption{a) Steady-state diffusion current per unit length per unit concentration and b) Settling time for the various 2-D nanogap redox cycling schemes shown in fig. \ref{2D_Electrode_Geometries} }
\label{2D_Result_1}
\end{figure}

The numerical simulation results for the electrode currents are plotted in Fig. \ref{2D_Result_1}-a. For scheme A, the electrode current increases with the electrode area. In the limiting case of planar electrodes, as expected,  the current is well predicted by the corresponding analytical results. However, we note that the increase in current is non-linear with the net electrode area. This is due to the strong 2D effects associated with diffusion-limited transport when electrodes are nanostructured. Interestingly, this result indicates that one can obtain near-ideal detection limits even when the total electrode area is small (e.g., the redox current is almost $90\%$ of the maximum even when the electrode area is only $50\%$). However, the associated statistics and variability could be a concern for detection at very low concentrations.

\vspace{0.7em}

The results for scheme B indicate a very similar trend (see Fig. \ref{2D_Result_1}-a). Here the total electrode area is kept the same, although the number of electrode pairs is progressively increased (see Fig. \ref{2D_Electrode_Geometries}). As a result, the 2D effects alone contribute to the observed increase in current from configuration $i$ to $iv$.

\vspace{0.7em}

\textbf{Nanostructured electrodes with planar layout}: Traditionally, nanogap electrodes are fabricated on the two sides of a channel, which requires complex fabrication process (see scheme A-$iv$ of Fig. \ref{2D_Electrode_Geometries}). Since the redox current increases inversely with the gap $W$, it is imperative to have a smaller gap for better detection limits - which have associated challenges concerned with the probability of molecules entering such a small gap. These contrasting optimization challenges can be effectively addressed through the scheme C of Fig. \ref{2D_Electrode_Geometries}.  Here, the electrodes are arranged on a planar surface. Accordingly, the target molecules could visit the electrodes without significant access issues. However, the diffusion of redox species is two dimensional, which could result in a reduction in electrode currents. Interestingly, the results in Fig. \ref{2D_Result_1}-a indicate that scheme C, with its planar architecture and reduced complexity in fabrication, shows strong scaling trends with nano-structuring. A two-dimensional nanostructuring (similar to a chequerboard pattern as opposed to a one-dimensional nanostructuring as in scheme C) is expected to yield more flux per unit concentration and hence could significantly improve the detection limits of redox cycling sensors.   

\vspace{0.7em}

\vspace{0.7em}

The results in Fig. \ref{2D_Result_1}-a indicates clear scaling trends with respect to nano-structured electrodes due to the influence of 2D diffusion.  However, an associated performance metric is the settling time to obtain steady-state results. Fig. \ref{2D_Result_1}-b shows the variation of settling time for the various schemes shown in Fig. \ref{2D_Electrode_Geometries}. Here, we find that 2D diffusion effects result in more settling time as compared to 1D diffusion. Accordingly, the conventional nanogap redox detection schemes perform better than the unconventional planar scheme in terms of settling time. The settling time for the sensors in the planar layout (scheme C) is about 45 $\mu$s, which is almost 3-4 times the settling time for the conventional parallel electrode scheme (12.4 $\mu$s).  The settling times approach the $W^2/D$ limit for schemes that resemble the planar electrode system (schemes A and B).
The scaling trends in Fig. \ref{2D_Result_1}-a,b indicates unconventionally patterned electrodes on a planar surface could yield better detection capabilities with reduced process complexity, although there could be a trade-off with respect to the settling time.

\section{Conclusions}
To summarize, here we proposed an electronic circuit inspired numerical simulation strategy for electrochemical biosensors. Through an insightful mapping of the diffusion problem to RC networks, we could reproduce well known analytical results related to 1D nanogap electrochemical biosensors. The same scheme, when extended to 2D geometry, allowed us to explore the effect of nanostructured electrodes. Our results indicate that nano-structuring results in enhanced 2D diffusion effects, which results in better area scaling trends, which could result in lower detection limits. However, the same results in increased settling times as well. The methodology proposed and insights revealed in this work could be of broad interest to a variety of applications related to electrochemical biosensors and storage cells as well.

\section{Conflict of Interest}
The authors declare no competing financial interest.

\bibliography{achemso}

\providecommand{\latin}[1]{#1}
\makeatletter
\providecommand{\doi}
  {\begingroup\let\do\@makeother\dospecials
  \catcode`\{=1 \catcode`\}=2 \doi@aux}
\providecommand{\doi@aux}[1]{\endgroup\texttt{#1}}
\makeatother
\providecommand*\mcitethebibliography{\thebibliography}
\csname @ifundefined\endcsname{endmcitethebibliography}
  {\let\endmcitethebibliography\endthebibliography}{}
\begin{mcitethebibliography}{24}
\providecommand*\natexlab[1]{#1}
\providecommand*\mciteSetBstSublistMode[1]{}
\providecommand*\mciteSetBstMaxWidthForm[2]{}
\providecommand*\mciteBstWouldAddEndPuncttrue
  {\def\EndOfBibitem{\unskip.}}
\providecommand*\mciteBstWouldAddEndPunctfalse
  {\let\EndOfBibitem\relax}
\providecommand*\mciteSetBstMidEndSepPunct[3]{}
\providecommand*\mciteSetBstSublistLabelBeginEnd[3]{}
\providecommand*\EndOfBibitem{}
\mciteSetBstSublistMode{f}
\mciteSetBstMaxWidthForm{subitem}{(\alph{mcitesubitemcount})}
\mciteSetBstSublistLabelBeginEnd
  {\mcitemaxwidthsubitemform\space}
  {\relax}
  {\relax}

\bibitem[Turner(2013)]{C3CS35528D}
Turner,~A. P.~F. Biosensors: sense and sensibility. \emph{Chem. Soc. Rev.}
  \textbf{2013}, \emph{42}, 3184--3196\relax
\mciteBstWouldAddEndPuncttrue
\mciteSetBstMidEndSepPunct{\mcitedefaultmidpunct}
{\mcitedefaultendpunct}{\mcitedefaultseppunct}\relax
\EndOfBibitem
\bibitem[Song \latin{et~al.}(2006)Song, Xu, and Fan]{Song2006}
Song,~S.; Xu,~H.; Fan,~C. Potential diagnostic applications of biosensors:
  current and future directions. \emph{International journal of nanomedicine}
  \textbf{2006}, \emph{1}, 433--440, 17722277[pmid]\relax
\mciteBstWouldAddEndPuncttrue
\mciteSetBstMidEndSepPunct{\mcitedefaultmidpunct}
{\mcitedefaultendpunct}{\mcitedefaultseppunct}\relax
\EndOfBibitem
\bibitem[Hashemi~Goradel \latin{et~al.}(2018)Hashemi~Goradel, Mirzaei,
  Sahebkar, Poursadeghiyan, Masoudifar, Malekshahi, and
  Negahdari]{doi:10.1002/jcb.26030}
Hashemi~Goradel,~N.; Mirzaei,~H.; Sahebkar,~A.; Poursadeghiyan,~M.;
  Masoudifar,~A.; Malekshahi,~Z.~V.; Negahdari,~B. Biosensors for the Detection
  of Environmental and Urban Pollutions. \emph{Journal of Cellular
  Biochemistry} \textbf{2018}, \emph{119}, 207--212\relax
\mciteBstWouldAddEndPuncttrue
\mciteSetBstMidEndSepPunct{\mcitedefaultmidpunct}
{\mcitedefaultendpunct}{\mcitedefaultseppunct}\relax
\EndOfBibitem
\bibitem[Yu \latin{et~al.}(2005)Yu, Blankert, Viré, and
  Kauffmann]{doi:10.1080/00032710500205659}
Yu,~D.; Blankert,~B.; Viré,~J.; Kauffmann,~J. Biosensors in Drug Discovery and
  Drug Analysis. \emph{Analytical Letters} \textbf{2005}, \emph{38},
  1687--1701\relax
\mciteBstWouldAddEndPuncttrue
\mciteSetBstMidEndSepPunct{\mcitedefaultmidpunct}
{\mcitedefaultendpunct}{\mcitedefaultseppunct}\relax
\EndOfBibitem
\bibitem[Turdean(2011)]{Turdean}
Turdean,~G.~L. {Design and Development of Biosensors for the Detection of Heavy
  Metal Toxicity}. \emph{{International Journal of Electrochemistry}}
  \textbf{2011}, \emph{2011}\relax
\mciteBstWouldAddEndPuncttrue
\mciteSetBstMidEndSepPunct{\mcitedefaultmidpunct}
{\mcitedefaultendpunct}{\mcitedefaultseppunct}\relax
\EndOfBibitem
\bibitem[Mohan \latin{et~al.}(2011)Mohan, Mach, Bercovici, Pan, Dhulipala,
  Wong, and Liao]{10.1371/journal.pone.0026846}
Mohan,~R.; Mach,~K.~E.; Bercovici,~M.; Pan,~Y.; Dhulipala,~L.; Wong,~P.~K.;
  Liao,~J.~C. Clinical Validation of Integrated Nucleic Acid and Protein
  Detection on an Electrochemical Biosensor Array for Urinary Tract Infection
  Diagnosis. \emph{PLOS ONE} \textbf{2011}, \emph{6}, 1--8\relax
\mciteBstWouldAddEndPuncttrue
\mciteSetBstMidEndSepPunct{\mcitedefaultmidpunct}
{\mcitedefaultendpunct}{\mcitedefaultseppunct}\relax
\EndOfBibitem
\bibitem[Lab(2011)]{Label}
{Label-Free Biosensors for Cell Biology}. \emph{{International Journal of
  Electrochemistry}} \textbf{2011}, \emph{2011}\relax
\mciteBstWouldAddEndPuncttrue
\mciteSetBstMidEndSepPunct{\mcitedefaultmidpunct}
{\mcitedefaultendpunct}{\mcitedefaultseppunct}\relax
\EndOfBibitem
\bibitem[Vercoutere and Akeson(2002)Vercoutere, and Akeson]{VERCOUTERE2002816}
Vercoutere,~W.; Akeson,~M. Biosensors for DNA sequence detection. \emph{Current
  Opinion in Chemical Biology} \textbf{2002}, \emph{6}, 816 -- 822\relax
\mciteBstWouldAddEndPuncttrue
\mciteSetBstMidEndSepPunct{\mcitedefaultmidpunct}
{\mcitedefaultendpunct}{\mcitedefaultseppunct}\relax
\EndOfBibitem
\bibitem[An \latin{et~al.}(2018)An, Wang, Han, Li, Jin, and Liu]{C7LC01117B}
An,~L.; Wang,~G.; Han,~Y.; Li,~T.; Jin,~P.; Liu,~S. Electrochemical biosensor
  for cancer cell detection based on a surface 3D micro-array. \emph{Lab Chip}
  \textbf{2018}, \emph{18}, 335--342\relax
\mciteBstWouldAddEndPuncttrue
\mciteSetBstMidEndSepPunct{\mcitedefaultmidpunct}
{\mcitedefaultendpunct}{\mcitedefaultseppunct}\relax
\EndOfBibitem
\bibitem[Yan and Fong(2016)Yan, and Fong]{Yan2016}
Yan,~Q.; Fong,~S.~S. In \emph{Systems Biology Application in Synthetic
  Biology}; Singh,~S., Ed.; Springer India: New Delhi, 2016; pp 53--70\relax
\mciteBstWouldAddEndPuncttrue
\mciteSetBstMidEndSepPunct{\mcitedefaultmidpunct}
{\mcitedefaultendpunct}{\mcitedefaultseppunct}\relax
\EndOfBibitem
\bibitem[Ronkainen \latin{et~al.}(2010)Ronkainen, Halsall, and
  Heineman]{B714449K}
Ronkainen,~N.~J.; Halsall,~H.~B.; Heineman,~W.~R. Electrochemical biosensors.
  \emph{Chem. Soc. Rev.} \textbf{2010}, \emph{39}, 1747--1763\relax
\mciteBstWouldAddEndPuncttrue
\mciteSetBstMidEndSepPunct{\mcitedefaultmidpunct}
{\mcitedefaultendpunct}{\mcitedefaultseppunct}\relax
\EndOfBibitem
\bibitem[Kang \latin{et~al.}(2013)Kang, Nieuwenhuis, Mathwig, Mampallil, and
  Lemay]{doi:10.1021/nn404440v}
Kang,~S.; Nieuwenhuis,~A.~F.; Mathwig,~K.; Mampallil,~D.; Lemay,~S.~G.
  Electrochemical Single-Molecule Detection in Aqueous Solution Using
  Self-Aligned Nanogap Transducers. \emph{ACS Nano} \textbf{2013}, \emph{7},
  10931--10937, PMID: 24279688\relax
\mciteBstWouldAddEndPuncttrue
\mciteSetBstMidEndSepPunct{\mcitedefaultmidpunct}
{\mcitedefaultendpunct}{\mcitedefaultseppunct}\relax
\EndOfBibitem
\bibitem[Chen \latin{et~al.}()Chen, Yousefi, Nemr, Gomis, Atwal, Labib,
  Sargent, and Kelley]{doi:10.1002/adfm.201907701}
Chen,~J.~B.; Yousefi,~H.; Nemr,~C.~R.; Gomis,~S.; Atwal,~R.; Labib,~M.;
  Sargent,~E.; Kelley,~S.~O. Nanostructured Architectures for Biomolecular
  Detection inside and outside the Cell. \emph{Advanced Functional Materials}
  \emph{n/a}, 1907701\relax
\mciteBstWouldAddEndPuncttrue
\mciteSetBstMidEndSepPunct{\mcitedefaultmidpunct}
{\mcitedefaultendpunct}{\mcitedefaultseppunct}\relax
\EndOfBibitem
\bibitem[Baur(2007)]{BAUR2007829}
Baur,~J.~E. In \emph{Handbook of Electrochemistry}; Zoski,~C.~G., Ed.;
  Elsevier: Amsterdam, 2007; pp 829 -- 848\relax
\mciteBstWouldAddEndPuncttrue
\mciteSetBstMidEndSepPunct{\mcitedefaultmidpunct}
{\mcitedefaultendpunct}{\mcitedefaultseppunct}\relax
\EndOfBibitem
\bibitem[Zafarani \latin{et~al.}(2016)Zafarani, Mathwig, Sudhölter, and
  Rassaei]{ZAFARANI201642}
Zafarani,~H.~R.; Mathwig,~K.; Sudhölter,~E.~J.; Rassaei,~L. Electrochemical
  redox cycling in a new nanogap sensor: Design and simulation. \emph{Journal
  of Electroanalytical Chemistry} \textbf{2016}, \emph{760}, 42 -- 47\relax
\mciteBstWouldAddEndPuncttrue
\mciteSetBstMidEndSepPunct{\mcitedefaultmidpunct}
{\mcitedefaultendpunct}{\mcitedefaultseppunct}\relax
\EndOfBibitem
\bibitem[Bard \latin{et~al.}(1986)Bard, Crayston, Kittlesen, Varco~Shea, and
  Wrighton]{doi:10.1021/ac00124a045}
Bard,~A.~J.; Crayston,~J.~A.; Kittlesen,~G.~P.; Varco~Shea,~T.; Wrighton,~M.~S.
  Digital simulation of the measured electrochemical response of reversible
  redox couples at microelectrode arrays: consequences arising from closely
  spaced ultramicroelectrodes. \emph{Analytical Chemistry} \textbf{1986},
  \emph{58}, 2321--2331\relax
\mciteBstWouldAddEndPuncttrue
\mciteSetBstMidEndSepPunct{\mcitedefaultmidpunct}
{\mcitedefaultendpunct}{\mcitedefaultseppunct}\relax
\EndOfBibitem
\bibitem[Madec \latin{et~al.}(2017)Madec, Lallement, and
  Haiech]{10.1371/journal.pone.0182385}
Madec,~M.; Lallement,~C.; Haiech,~J. Modeling and simulation of biological
  systems using SPICE language. \emph{PLOS ONE} \textbf{2017}, \emph{12},
  1--21\relax
\mciteBstWouldAddEndPuncttrue
\mciteSetBstMidEndSepPunct{\mcitedefaultmidpunct}
{\mcitedefaultendpunct}{\mcitedefaultseppunct}\relax
\EndOfBibitem
\bibitem[Bard and Faulkner(1980)Bard, and Faulkner]{Bard1980ElectrochemicalMF}
Bard,~A.~J.; Faulkner,~L.~R. Electrochemical Methods: Fundamentals and
  Applications. 1980\relax
\mciteBstWouldAddEndPuncttrue
\mciteSetBstMidEndSepPunct{\mcitedefaultmidpunct}
{\mcitedefaultendpunct}{\mcitedefaultseppunct}\relax
\EndOfBibitem
\bibitem[Berg and Purcell(1977)Berg, and Purcell]{BERG1977193}
Berg,~H.; Purcell,~E. Physics of chemoreception. \emph{Biophysical Journal}
  \textbf{1977}, \emph{20}, 193 -- 219\relax
\mciteBstWouldAddEndPuncttrue
\mciteSetBstMidEndSepPunct{\mcitedefaultmidpunct}
{\mcitedefaultendpunct}{\mcitedefaultseppunct}\relax
\EndOfBibitem
\bibitem[Causon \latin{et~al.}(2011)Causon, Mingham, and Qian]{book}
Causon,~D.; Mingham,~C.; Qian,~L. \emph{Introductory finite volume methods for
  PDEs}; 2011\relax
\mciteBstWouldAddEndPuncttrue
\mciteSetBstMidEndSepPunct{\mcitedefaultmidpunct}
{\mcitedefaultendpunct}{\mcitedefaultseppunct}\relax
\EndOfBibitem
\bibitem[Pierret and Harutunian(1996)Pierret, and
  Harutunian]{pierret1996semiconductor}
Pierret,~R.; Harutunian,~K. \emph{Semiconductor Device Fundamentals};
  Addison-Wesley, 1996\relax
\mciteBstWouldAddEndPuncttrue
\mciteSetBstMidEndSepPunct{\mcitedefaultmidpunct}
{\mcitedefaultendpunct}{\mcitedefaultseppunct}\relax
\EndOfBibitem
\bibitem[{Madec} \latin{et~al.}(2019){Madec}, {Hébrard}, {Kammerer},
  {Bonament}, {Rosati}, and {Lallement}]{8602461}
{Madec},~M.; {Hébrard},~L.; {Kammerer},~J.; {Bonament},~A.; {Rosati},~E.;
  {Lallement},~C. Multiphysics Simulation of Biosensors Involving 3D Biological
  Reaction–Diffusion Phenomena in a Standard Circuit EDA Environment.
  \emph{IEEE Transactions on Circuits and Systems I: Regular Papers}
  \textbf{2019}, \emph{66}, 2188--2197\relax
\mciteBstWouldAddEndPuncttrue
\mciteSetBstMidEndSepPunct{\mcitedefaultmidpunct}
{\mcitedefaultendpunct}{\mcitedefaultseppunct}\relax
\EndOfBibitem
\bibitem[Nair and Alam(2013)Nair, and Alam]{C2AN35346F}
Nair,~P.~R.; Alam,~M.~A. A compact analytical formalism for current transients
  in electrochemical systems. \emph{Analyst} \textbf{2013}, \emph{138},
  525--538\relax
\mciteBstWouldAddEndPuncttrue
\mciteSetBstMidEndSepPunct{\mcitedefaultmidpunct}
{\mcitedefaultendpunct}{\mcitedefaultseppunct}\relax
\EndOfBibitem
\end{mcitethebibliography}

\end{document}